# Mastering energy landscapes *via* liquid-liquid phase separation to program active supramolecular co-assembly from the nano to macro scale

Yuanhao Wu[1,2,3,4]\*, Alexander van Teijlingen[5,6], Julie Watts[4,7], Zhiquan Yu[1,2], Shanshan Su[1,2], José Carlos Rodríguez-Cabello[8], Lihi Abramovich[9], Tell Tuttle[6], Alvaro Mata[3,4,10]\*

1. Plastic and Reconstructive Surgery Department, Wuhan Union Hospital, Tongji Medical College, Huazhong University of Science and Technology, Wuhan, China

2. Plastic and Reconstructive Surgery Research Institute, Wuhan Union Hospital, Tongji Medical College, Huazhong University of Science and Technology, Wuhan, China

3. Biodiscovery Institute, University of Nottingham, NG7 2RD Nottingham, UK

4. School of Pharmacy, University of Nottingham, NG7 2RD Nottingham, UK

5. ModelMole, Glasgow, G52 1NB, UK

6. Department of Chemistry, University of Strathclyde, Glasgow G1 1XL, UK

7. Nanoscale and Macroscale Research Centre, University of Nottingham, NG7 2RD Nottingham, UK

8. BIOFORGE Lab, University of Valladolid, LaDIS, CIBER-BBN, 47011 Valladolid, Spain

9. Department of Oral Biology, The Goldschleger School of Dental Medicine, Sackler Faculty of Medicine, Tel Aviv University, Tel Aviv 6997801, Israel

10. Department of Chemical and Environmental Engineering, University of Nottingham, Nottingham NG7 2RD, UK

To whom correspondence should be addressed.
E-mails: Yuanhaowu@hust.edu.cn; A.Mata@nottingham.ac.uk

**Abstract**

The energy landscape dictates pathways and outcomes in supramolecular self-assembly, yet harnessing it from the nano to the macro scales remains a major challenge. Here, we demonstrate liquid–liquid phase separation (LLPS) as a powerful tool to navigate and engineer the energy landscapes of co-assembly systems comprising disordered proteins and peptides. We quantitatively map the energy barriers and transition states governing structural transitions, enabling predictive "on/off" control of assembly and hierarchical order from nano to macro scales. By integrating supramolecular biofabrication, we achieve spatially organized architectures with life-like non-equilibrium behaviour. Crucially, assembly stability and scalable self-sorting are shown to depend on accessing minimum-energy states, regardless of whether the co-assembled structures are disordered or ordered. This work establishes energy landscape mediation *via* LLPS as a general framework for designing life-like, hierarchically structured materials.

## Introduction

Energy landscapes regulate interactions[1–3], nano and microstructures[4,5], and mechanical properties[6–8] of supramolecular self-assembling materials[9]. Multi-component self-assembling (co-assembling) systems composed of different molecular building-blocks result in more complex materials given the extended spectrum of potential interactions[10–16], concentration ratios[17], active sites[18,19], and orders of reaction[20–23]. By regulating these parameters, it is possible to tune the specific assembling mechanisms, structures, and corresponding properties, opening opportunities to rationally design more complex supramolecular materials[24]. One way to describe the resulting structures and properties of these materials is through intricate energy landscapes, which for example can be used to define equilibrium and non-equilibrium self-assembling systems[25–27]. The energy landscapes are difficult to predict and investigate due to the diverse pathways to access each energy state for specific microstructures[28,29]. The capacity to predict and harness the energy landscapes of co-assembling systems would open opportunities to rationally design materials with tuneable structural hierarchy and non-equilibrium properties leading to responsive functions of living systems.

LLPS has been used to study the relationship between energy landscapes and physical properties of corresponding co-assembling structures. LLPS is an energetically favourable process[30,31] that is determined by a) the concentration of macromolecules in aqueous solutions, b) the nature of the macromolecules (length, hydrophobicity, charge distribution), and c) the surrounding biophysical conditions (temperature, pH, ionic concentration). An example of LLPS in biology is the formation of membraneless organelles (MLOs)[32–35], which are biomolecular condensates such as nucleoli[36], Cajal bodies[37], and nuclear speckles present in cells with functions such as the regulation of gene expression. Through these LLPS mechanisms[38], it may be possible to design new co-assembling structures (e.g. fibrils[39], tactoids[40], micelles[41]) bridging energy landscapes with innovative desirable functional properties.

Proteins exhibit intrinsically disordered regions (IDRs), which can undergo disorder-to-order transitions that drive and regulate LLPSs[32,42,43]. For example, MLOs are rich in proteins containing IDRs which interact amongst them to form MLOs with different properties such as gene-regulatory roles in transcription or RNA processing and translation[44–46]. In artificial systems, proteins comprising IDRs have also been used in combination with LLPS processes to engineer hierarchical structures. For example, a viscoelastic chromatin tethering and organization mechanism using LLPS-based biomolecular condensates driven by IDR-IDR interactions has been reported to generate capillary forces at targeted DNAloci[47,48]. We have also demonstrated how LLPS can be combined with compartmentalization or diffusion-reaction processes to tune disorder-to-order transitions and assemble IDR-containing proteins hierarchically into macroscopic structures such as tubes[49] and membranes[50]. These systems can also be integrated with biofabrication processes to further control hierarchical assembly in a reproducible manner[51]. While these systems offer unique opportunities to introduce bioactive molecularly-designed nanostructures within the field of biofabrication, the fundamental mechanisms underlying the supramolecular design of disordered and ordered components remain poorly understood.

To realize this supramolecular biofabrication vision and how they relate with disorder and order structures, it is critical to better understand the energy landscape driving supramolecular co-assembly and more accurately predict the assembly of structures defined by LLPS. Achieving these goals would not only help bridge current knowledge gaps in LLPS of multicomponent systems, but also enable LLPS to be used to rationally design co-assembling material systems with hierarchical precision from the nano to the macroscale. To achieve this, the energy landscape would serve as a design toolkit to unify multicomponent self-assembly across size-scales, integrating co-assembly at the supramolecular-to-nanoscale, LLPS at the nano-to-microscale, and printing at the micro-to-macroscale.

Here, we report on a disordered protein / self-assembling peptide non-equilibrium co-assembling system, which enables the hierarchical control of molecular interaction at solution interface *via* LLPS. We demonstrate how the approach can predict "on" and "off" states and navigate between, as well as assembling nano- and micro-structures, which can be further assembled at higher size-scales by printing. Furthermore, by harnessing the energy landscape of these structures, it is possible to make with quantitively calculations of the energy barrier for each structural pathway. This LLPS-based system bridges nano to macro control by using transition states of material energy landscapes to predict molecular interactions and hierarchical co-assembly. Finally, we demonstrate it does not matter the components of disordered or ordered structures in the co-assembly hybrids, rather, achieving a structure at the minima of energy landscape is critical to system stability and potential to self-sorting from nano- to macro-scale spontaneously with life-like non-equilibrium behaviour.

**Results**
**Rationale of the design**
We used an LLPS system based on the co-assembly of an elastin-like recombinamer (ELK1) as a model of a disordered protein and the Fmoc-based self-assembling peptide FmocFFK (FK) (Figure 1a and 1b). These molecules were selected because of their unique energy landscape profiles, which can be described based on their transition state (TS) by indicating the minimum energy gap [52]. ELK1 exhibits a temperature-dependent TS (at its transition temperature, $T_t$), as previously reported [53], while FK displays a pH-dependent TS. FmocFFR (FR) was used as control because of its similar molecular structure to FK but lacking a TS (Figure 1a). Experiments were conducted at different pH values selected to demonstrate the existence of TS in the self-assembly of FK and describe its energy landscape. Subsequently, we assessed the co-assembly of ELK1-FK into multiple supramolecular conformations (Figure 1c) exhibiting different energy states (Figure 1d-f) to help describe their energy landscapes. Finally, Laponite® was introduced into a solution of FK-ELK1 to trigger self-sorting by increasing the disorder of system and lowering the energy states of the different resulting structures (Figure 1g). *In situ* imaging, physicochemical calculations, and coarse-grained molecular dynamics (CGMD) computational simulations were conducted to describe the underlying assembly mechanism while fabrication methods were used to demonstrate control and applicability of the system.

**The hierarchical structure and energy landscape of FK**
*Self-assembled nanostructures of FK at different pH values and its TS at pH5*
To describe the energy landscape of the peptide amphiphiles FK, we first investigated the role of pH in its self-assembly. Fmoc-based peptides are known to be sensitive to pH, resulting in the self-assembly of different nanostructures [52]. Therefore, we first used Cryo-TEM to describe the resulting self-assembled structures of FK at pH3, 5, and 7. The results revealed that FK self-assembled into ribbons at pH3, twist ribbons at pH5, and bundled ribbons at pH7 (Figure 2a). To investigate the intermolecular interactions behind these self-assemblies, we used CGMD (Figure 2b), which revealed that hydrophobic aggregation was regulated predominantly by the hydrophilic and electrostatic interactions of the lysine (Lys) residues and π-π stacking of the Fmoc moiety. To a lesser extent, π-π interactions with and between Phenylalanine (Phe) residues were also present. The simulation results revealed that self-assembled 3D nanostructures were formed at pH3–7 and had the highest solvent accessible surface area (SASA) at pH5 (Figure 2c and Supplementary Figure 2). The order of SASA created a discontinuity in the correlation between surface charge and surface area. One would expect SASA to be ordered the same as deprotonation, that is pH7 > 5 > 3. However, we observed a SASA order following pH5 > 7 > 3 (Supplementary Figure 3). From this finding, it could be determined that the Phe residue of FK was not the primary driver of the hydrophobic interactions as it is often presented at the surface of the aggregate. Instead, the hydrophobic interaction of FK was primarily driven by the Fmoc protecting group. FK self-assembled structures at pH5 exhibited a higher SASA than either of those formed at pH3 or pH7 (Figure 2c); thus, FK at pH5 could be thought of as a kind of TS between the other two states.

*The intermolecular antiparallel interlock of Fmoc-Fmoc leading to FK's TS at pH5*
To further understand the TS of FK at pH5, CGMD was used to investigate intermolecular interactions between neighbouring FK molecules. At pH3–7, the backbone (BB) dihedrals for peptides containing a protonated Lys were clustered around 1 & -1 *cos* (◦) with high variability in between (Supplementary Table 1). At the location where the Lys was deprotonated, the backbone dihedrals shift towards 1 *cos* (◦) (representing a curling up from a more extended conformation), though the magnitude of this effect decreased with the number of deprotonated Lys (the effect is strongest at pH5). A lower *cosine* dihedral angle between Phe residues within FK indicated that Phe residues were less involved in the hydrophobically driven self-assembly process.[54] When Lys became deprotonated, the Phe-BB-BB-Phe dihedral shifted higher, which indicates intramolecular stacking, an effect that is strongest at pH5 (Supplementary Section 1). To further understand the intermolecular interaction of TS by looking at the simulation results of neighbour FK molecules, the stacking around Fmoc units were observed to be typically parallel. The neighbour distance (minimum Fmoc-Fmoc centre of geometry distance) of each FK monomer was defined by measuring the BB bead distance from the two Fmoc groups. We found that, at pH5, the selectivity for antiparallel stacking was higher than at either pH3 or pH7 (Figure 2d). This result was consistent with experimental circular dichroism (CD) spectroscopy (Figure 2e) and fluorescence emission (Figure 2f) data, which indicated a decrease of β-sheet components of FK at pH5 (Figure 2e) due to the antiparallel interlock of Fmoc-Fmoc (Figure 2f) stacking. This minor intermolecular interaction change led to the TS of FK self-assembly at pH5. These results demonstrate that the TS of FK self-assembly at pH5 is based on a nanostructure property of intermolecular antiparallel interlocking of Fmoc-Fmoc.

*The energy landscape of FK related with pH*
Having demonstrated the existence of the TS of FK at pH5 and the intermolecular interaction behind its self-assembly, we then investigated the energy landscape of FK self-assembly by using dynamic light scattering (DLS) to calculate the energy gap between FK states at different pHs (Figure 2g, Supplementary Section 1). The results revealed that the energy gap of FK transferring from pH3 to pH5 ($\Delta E_{3\rightarrow 5}$) was 33.65 kJ/mole, while $\Delta E_{5\rightarrow 7}$ and $\Delta E_{7\rightarrow 5}$ were 27.61 kJ/mole and 53.46 kJ/mole, respectively. Noteworthy, the transition of FK from pH5 to pH3 was too fast to track and analyse quantitatively, which meant that $\Delta E_{5\rightarrow 3}$ was the smallest among all these $\Delta E$ values (Figure 2g). Our experiments also found that the self-assembled FK structures became amorphous aggregates at pH 10, which was the most stable state among all pH conditions (Supplementary Section 1 and Figure 1c). These $\Delta E$ described the energy barrier to overcome from one structure state to another, which produced an energy landscape profile of FK self-assembly by indicating 2 quantitative pathways from the TS of FK (pH5) to its more stable states at pH3 and pH7 (Figure 2g). These results revealed that the energy barriers of the pathways $\Delta E_{5\rightarrow 3}$ and $\Delta E_{5\rightarrow 7}$ were easier to overcome compared to those of pathways $\Delta E_{3\rightarrow 5}$ and $\Delta E_{7\rightarrow 5}$ (Figure 2g). However, only the pathway $\Delta E_{5\rightarrow 7}$ corresponded to a self-assembly process while pathways $\Delta E_{5\rightarrow 3}$ were associated to dis-assembling processes generating less ordered structures. Therefore, $\Delta E_{5\rightarrow 7}$ was observed as the most promising state change to harness and generate rapid phase separations in water or through a liquid-liquid interface by co-assembling, such as with ELK1.

## The co-assembly of FK-ELK1
*FK-ELK1 intermolecular interaction*
To understand the interfacial LLPS of FK-ELK1, we first investigated the co-assembly between FK and ELK1. We used ELK1, at its original pH (pH5) and below Tt, as a chaperone to tune the structures of FK by directly mixing two solutions of each component (Figure 3a) and further co-assembling them at their solution interface (Figure 3f) at different pHs. Cryo-TEM observations revealed that the co-assembled FK-ELK1 fibrous nanostructures (Figure 3a) exhibited different morphologies. Interestingly, it was possible to tune the co-assembled nanostructures of FK-ELK1 into nanostructures that were similar to those of self-assembled FK at the next higher pH. In other words, the co-assembled nanostructures of FK (pH3)-ELK1 (Figure 3a ①) exhibited a similar twisted ribbon as those of FK at pH5 (Figure 2a ②). The co-assembling molecular interactions between ELK1 and FK were analyzed using CGMD (Figure 3b and Supplementary Section 2 and 3) and demonstrated that ELK1 was able to tune the self-assembled nanostructure of FK depending on the pH used. Similar nanostructures were formed upon co-assembly at adjacent pH values [*i.e.* FK (pH3)-ELK1 = FK (pH5), FK (pH5)-ELK1 = FK (pH7), FK (pH7)-ELK1 = FK (pH10)] through hydrophobic interactions taking place at the Phe residues of the FK. The simulations also revealed that the Fmoc groups were buried in the self-assembled FK nanostructures and did not contribute to the FK-ELK1 co-assembly, but helped to maintain the main networks of nanostructures with tiny antiparallel switch (Figure 3c, d, e, and Supplementary Figure 9). In conclusion, the featured fibres of FK-ELK1 or FK nanostructures were similarly affected by either conditions, spatial adaptations of ELK1 chaperones or pH changes, both of which enabled using a similar energy landscape to describe both FK self-assembly and FK-ELK1 co-assembly (Figure 2g).

*Nano- to macro-scale structural control via LLPS using the TS of FK at pH5*
To investigate the interaction between ELK1 proteins and FK at a fixed pH (pH5), we observed that the stacking structure of FK underwent a transformation, accompanied by a reduction in the relative amount of antiparallel stacking at the transition state (TS) of FK (Figure 3c). This was supported by a decrease in β-sheet content, as evidenced by circular dichroism (CD) (Figure 3d) and a blue shift in the fluorescence emission spectrum, centered around 330 nm (Figure 3e), after co-assembly at pH5. At pH5, compared to pH3 and pH7, the self-assembled FK structures exhibited the highest SASA and were enriched with Phe residues, which facilitated hydrophobic interactions with ELK1. CGMD further revealed that self-assembled FK structures retained a significant (~99%) proportion of protonated Lys residues at the surface, promoting electrostatically driven reorganization of ELK1 (Figure 3b and Supplementary Figure 3). Both the hydrophobic interactions of Phe and the electrostatic interactions of Lys played key roles in the formation of stable FK (pH5)-ELK1 nanostructures, with the most robust hierarchical assemblies were observed at this pH. These structures reached their minimal energy state of the energy landscape (Figure 2g), enabling the rapid formation of stable FK-ELK1 macroscopic structures through LLPS (Figure 3f). A filamentous macrostructure was formed by injecting FK (pH 5) into an ELK1 solution (Figure 3e②), as confirmed by confocal microscopy. No comparable structures were observed in systems with FK (pH 3)-ELK1 (Figure 3e①) or FK (pH 7)-ELK1 (Figure 3e③). These results demonstrate that by manipulating the nanostructure of FK-ELK1 co-assemblies at the TS, with minimal energy barriers, it is possible to drive the system towards a low-energy, ordered structure, providing a strategy for controlling hierarchical structures from the nanoscale to the macroscale.

**The spatiotemporal control and dynamic properties across nano- to macro- scales**
*The concentration ratios of FK-ELK1 co-assembly*
A typical macrostructure, such as a hydrogel, relies on molecular networks to serve as scaffolds that maintain significant amounts of water. The co-assembled macroscale structure, robust or viscous hydrogel, is determined by the dominant nanostructures fibres (self-assembled FK) or micelles (self-assembled ELK1) in this system. When FK fibrillar nanostructures function as dominant networks, encapsulating ELK1 micelles (Figure 3 and Supplementary Section 2). Alternatively, ELK1 micellar nanostructures dominate the whole system. Thus, a specific concentration ratio of ELK1 to FK was needed to be determined to have no excess components and further understand the dominant nanostructures. We used a titration approach harnessing the Tt of ELK1 by varying FK concentrations from 0.1% to 1%, while maintaining ELK1 at a fixed concentration of 2%. The choice of 2% ELK1 was based on its significant impact on the transition temperature (Tt) [53] (Figure 4a). ELK1 alone self-assembled into micelles at and above Tt (Supplementary Figure 11). When ELK1 excess, upon adding FK, the Tt of the FK-ELK1 system increased linearly (Figure 4b), as observed from UV-Vis spectroscopy, showing sharp slopes at Tt. As FK occupied ELK1 and restricted its collapse, a higher Tt was required to trigger aggregation of the excess ELK1. Once no excess components, the Tt of the FK-ELK1 system would disappear (Figure 4a). By fitting a linear relationship between FK concentration and Tt, we identified the optimal ratio of FK:ELK1 with no excess components (0.55%: 2%). These findings suggest the formation of three possible co-assembly conditions: (1) no excess FK (0.55% FK), (2) excess ELK1 (<0.55% FK), and (3) excess FK (>0.55% FK).

*Tuning FK-ELK1 co-assembly at nano- and macro-scale using TS of ELK1 (Tt) and FK (pH)*
Having determined the optimal FK-ELK1 concentration ratio without excess components, we then explored the hierarchical manipulation of the co-assembled structures by controlling the self-assembly of excess components. Initially, we selected a 0.2% FK (green) - 2% ELK1 (purple) system, where co-assembly was observed using confocal microscopy (Figure 4c, left). The co-assembled structure was triggered by the transition temperature (Tt) of the system, which was identified *via* UV-Vis spectroscopy (Figure 4b, ~37°C). Despite the presence of excess FK, ELK1 molecules continued to self-assemble into core-shell micelles (FK core, ELK1 shell), transforming the system into a viscous gel that could be reversibly transition by changes in temperature (Figure 4c, right). This reversible behaviour mirrored that of self-assembled ELK1, where excess ELK1 acted as a scaffold without disrupting the FK-ELK1 co-assembly. This system was reversibly tuned between a liquid and micellar viscous gel at the macroscale. A similar phenomenon was observed in the FK-excessive system (0.7% FK - 2% ELK1), where excess FK was triggered by pH and self-assembled into co-assembled ribbons. FK formed a scaffold network to which ELK1 molecules adhered (Figure 4d). This system also exhibited reversible transitions between a viscous gel and a suspension at the macroscale.

In conclusion, the order of reaction and concentration ratios were critical for dynamically controlling hierarchical co-assembling structures of FK and ELK1. Excess FK or ELK1 self-assembled into higher-energy states, requiring additional energy input provided *via* temperature or pH adjustment, whereas co-assembly produced more stable nanostructures due to stronger intermolecular forces. Through carefully tuning the order of reaction and the concentration ratios, it was possible to control the self-assembly process without disrupting the co-assembly structure, enabling dynamic control of hierarchical structures step-by-step.

*Reaching the minimum energy state by inducing disordered components*
To further validate this approach, we introduced a third component, a nanoclay hydrogel, Laponite® (L), into the FK-ELK1 co-assembly system. Laponite®, a negatively charged material, was chosen due to its electrostatic interaction with the positively charged FK and ELK1, as confirmed by zeta potential measurements (Figure 5a). These electrostatic interactions facilitated the co-assembly of Laponite® with FK or/and ELK1, resulting in a more robust nanostructure. DLS analysis revealed the formation of significant aggregates, including ELK1-Laponite® (EL), FK-Laponite® (FL), and ELK1-FK-Laponite® (EFL) complexes (Figure 5b). Given electrostatic forces were stronger than the hydrophobic interactions between FK and ELK1, Laponite® was able to disrupt the FK-ELK1 co-assembly. This result was confirmed by Cryo-TEM images, which verified that the original self-assembled structures of FK, ELK1, and FK-ELK1 were disrupted upon Laponite® incorporation (Figure 5c). SEM images revealed distinct fibrillar networks in EFL, EL, and FL hydrogels (Figure 5d). These disordered nanostructures corresponded to the minimum energy state of the system, indicating the most stable configuration of the system [55].

*Self-sorting and no-equilibrium properties by generating disordered nanostructures at the energy landscape minimum via LLPS*
To exploit the stable state of the co-assembled systems, we applied interfacial LLPS to test their printability using liquid-in-liquid printing. Following injection of aqueous Laponite® suspensions into FK, ELK1, or FK-ELK1 solutions, robust gel-like membranes were formed at the interfaces, which generated tubular structures at macroscale as shown by stereomicroscopy (Figure 6e). This phase separation enabled liquid-in-liquid printing, demonstrating that even disordered nanostructures could be utilized for printing when operating at the energy landscape minimum of each system. We further explored self-sorting at the liquid-liquid interface by injecting Laponite® into three FK-ELK1 systems with varying concentrations including 0.5% FK (ELK1-excessive), 0.55% FK (no excess), and 0.6% FK (FK-excessive), and 2% ELK1. As expected, self-sorting took place at the solution interfaces. In the "no excess" system, the ELK1 (purple) and FK (green) interfaces overlapped, with no excess components interacting with Laponite®. In the "FK-excessive" system, the FK-Laponite® (green) interface aligned closer to the Laponite® side, while the "ELK1-excessive" system showed the opposite configuration (Figure 5e), as confirmed by confocal microscopy. SEM images further validated these observations (Figure 5e). Additionally, reversible purple colour changes (homogeneous - aggregate -homogeneous) were observed inside the EL membrane microstructures were observed in the ELK1-excessive system, where temperature-induced self-assembly was partially restricted by Laponite® but remained responsive to temperature recycling changes (below Tt - above Tt - below Tt) (Figure 5f). In summary, strong intermolecular forces, such as electrostatic interactions, overrode weaker forces, facilitating stable self-sorting membranes and demonstrating life-like non-equilibrium behaviour. This process was achieved by generating nanostructures at the minima of the energy landscape through liquid-liquid phase separation, even when the nanostructures were disordered aggregates.

*Diversity of nano-, micro-, and macro-structures, mechanical properties, and biofunctions*
The nanostructures at the energy landscape minimum—whether ordered (FK pH 5) or disordered (EL/FL/EFL)—enabled the generation of stable macroscopic structures through interfacial LLPS. Rapid co-assembly of different systems (FK and ELK1, EF and L) at the solution interface facilitated liquid-in-liquid printing. However, the different nanostructure properties, such as porosity and diffusion directionality, contributed to a diversity of macroscale structures. When FK (pH 5) was injected into ELK1, a filamentous macrostructure was formed (Figure 6a), exhibiting a homogeneous gel-like structure as shown by birefringence (Figure 6a, inset), and high fidelity as confirmed by SEM (Figure 6b). This system enabled 3D printing of multilayered grids (Figure 6c) and exhibited high biocompatibility with human umbilical vein endothelial cells (hUVECs) (Figure 6d). The FK (pH 5)-ELK1 nanostructures contained longer fibers and larger pores (Figure 6b) than the EFL system (Figure 6f), which facilitated the formation of tubular macrostructures due to reduced molecular penetration through the interfacial barrier, resulting in a self-sorting membrane (Figure 6g). In another application, the ELK1-excess-FK micelle gels (0.2% FK - 2% ELK1) exhibited a Tt of 37°C, where FK molecules attached to suspended cells, and excess ELK1 self-assembled into gels upon incubation (Figure 6h). This system provided a 3D environment that promoted tube formation by hUVECs, even without growth factors (Figure 6i). These results demonstrate the hierarchical control of diverse nano-, micro-, and macroscale structures, mechanical properties, and potential biological functions in co-assembly systems by leveraging liquid-liquid interfacial phase separation and energy landscape manipulation.

**Discussion**
Our work establishes LLPS as a robust method to engineer the energy landscape of supramolecular co-assembly systems. With the application of quantitative mapping of the energy barriers and transition states that govern assembly pathways, we achieved predictive control over hierarchical organization from the nanoscale to the macroscale. We demonstrate the systems can be switched between "on and off" states and, critically, navigated across energy minima- irrespective of whether the co-assembled structures are ordered or disordered. These findings highlight energy landscape navigation as a central design principle for stable, non-equilibrium materials. Furthermore, this approach bridges traditionally distinct scales and states of matter through a unified energy-driven mechanism, enabling integration with fabrication techniques such as 3D printing. While challenges remain in resolving increasingly complex multicomponent landscapes, our strategy provides a general framework for designing life-like systems with adaptive and programmable properties. Ultimately, mastering energy landscapes *via* LLPS opens new avenues in biomaterials and synthetic biology, facilitating the bottom-up creation of intelligent, self-organizing matter.

**Methods**

**Materials**
FK (purity, 99.%), FR (purity, 98.%) were designed as reported previously [52,56] and purchased from GL BioChem (Shanghai, China). Laponite® (682659) was ordered from Sigma-Aldrich.

**Cryo-TEM**
Cryo-transmission electron microscopy (cryo-TEM) was performed to examine the sample morphology. Solutions of FK and ELK1 were prepared at a concentration of 0.1 wt% in Milli-Q water. Quantifoil R 1.2/1.3 grids were glow-discharged in air for 60 s at 40 mA using a Quorum GloQube® system. Prior to vitrification, grid tweezers, grids, and pipette tips were preheated to the sample temperature to prevent aggregation. A 3μL aliquot of each sample was applied to the grid, followed by blotting for 5-6 s with a blot force of 6 using a Vitrobot MK IV (Thermo Fisher Scientific). Vitrified grids were imaged under a FEI Tecnai G12 12 Biotwin microscope (Thermo Fisher Scientific) operating at 200 kV, equipped with a HADDF detector.

**Zeta potential ($\zeta$)**
To characterize the co-assembly behaviour among ELK1, FK, and Laponite®, $\zeta$ potential measurements were performed using a Zetasizer Nano-ZS instrument (Malvern Instruments, Worcestershire, UK). All experiments were conducted at room temperature using solutions at a concentration of 0.1 wt%. The pH of each solution was adjusted to match the corresponding pH of the 2 wt% PA solutions, using 1 M HCl or 1 M $NH_4OH$ as required. Samples were equilibrated for 30 minutes at room temperature before measurement.

**Circular dichroism (CD)**
CD spectra were acquired using a Chirascan™ CD spectrometer (Applied Photophysics, UK) at a temperature of 25 °C. Samples of ELK1 (0.01 wt%) and FK (0.01 wt%) were prepared in Milli-Q water and equilibrated at room temperature for 10 minutes prior to data collection. Measurements were performed in a quartz cuvette with a 0.1 cm path length. Spectra were recorded from 190 to 260 nm at a scan speed of 50 nm/min, with each spectrum representing the average of 10 accumulations. Raw data were processed using a moving-average smoothing algorithm.

**Fluorescence Spectroscopy**
Fluorimetric analyses were conducted using a Perkin-Elmer LS55 luminescence spectrometer equipped with a temperature-controlled unit (Julabo F25). Samples of ELK1 (0.01 wt%) and FK (0.01 wt%) were prepared in Milli-Q water and equilibrated at room temperature for 10 minutes prior to data collection. Following preparation, samples were transferred into 1 cm pathlength disposable PMMA cuvettes (Fisher Scientific). All measurements were performed at 25 °C. Emission spectra were recorded across a wavelength range of 300–600 nm upon excitation at 265 nm, using slit widths of 10 nm (excitation) and 3 nm (emission), and a scan speed of 300 nm min$^{-1}$. Data acquisition was carried out using the FL WinLab software. Variations in pH conditions induce changes in the self-assembled structures, which may alter the sample refractive index and introduce quenching effects due to molecular packing. These factors can influence the absolute fluorescence intensity by affecting light transmission and quenching related to self-assembly. To enable meaningful comparison across samples, all emission spectra were normalized to the intensity of their respective maximum peak.

**Calculation of energy barrier (ΔE)**
The fraction of self-assembled FK (α), was determined from time-dependent static light-scattering (SLS) intensity measurements, which were performed using a Linkoptik Nanolink SZ9011 instrument (Linkoptik, China) [57]. A pH-induced assembly process was initiated by adding 20 µL of either NaOH or HCl aqueous solution to 0.01 wt% aqueous solutions of FK. Data processing and Arrhenius fitting were conducted in accordance with established methods [58].

$$\alpha = \frac{\text{scattering intensity at a given time} - \text{scattering intensity of the product}}{\text{scattering intensity of the reagent} - \text{scattering intensity of the product}} \quad \ldots\ldots (1)$$

In conjunction with equation (2) described by Wilkinson [59], the alpha value as a function of time can be obtained, where $k$ is a rate constant, n is the reaction order, $t$ is the time, $c_0$ the initial concentration of FK.

$$\alpha = \frac{2 + kc_0(n-2)t}{2 + kc_0 nt} \quad \ldots\ldots (2)$$

Fitting the rate constant $k$ and temperature T (absolute temperature) to Arrhenius equation (1) yields energy barrier values ΔE for each reaction.

**UV–visible spectrophotometer to monitor the Tt of ELK1-FK systems**
The thermo-responsive behavior of ELK1 at a concentration of 2 wt% and FK at varying concentrations from 0 wt% to 0.6 wt% were characterized using a temperature-controlled UV–visible spectrophotometer (Agilent Technologies). Sample solutions were prepared in Milli-Q water, and the pH was adjusted using 0.5 M HCl and 1.0 M NH$_4$OH. The temperature was increased at a constant rate of 1 °C/min while monitoring the absorbance at λ = 350 nm.

**Confocal microscopy**
The interaction and co-localization of ELK1 and FK were examined using laser scanning confocal and multiphoton microscopy (TCS SP2, Leica Microsystems, Germany). ELK1 (2 wt%) was dissolved in an aqueous solution containing Alexa Fluor™ 647 NHS Ester ($10^{-6}$ wt%) (Thermo Fisher Scientific), while FK was diluted to varying concentrations in an aqueous solution of Thioflavin T (ThT) ($10^{-6}$ wt%) (596200, Sigma-Aldrich). All solutions were incubated for 20 minutes at designed temperatures in the dark. For imaging, Image acquisition was performed using laser excitation wavelengths at 647 nm and 450 nm, corresponding to the optimal peaks of Alexa Fluor 647 and ThT, respectively. Subsequent image processing and analysis were conducted using ImageJ software. *In situ* temperature was adjusted by a temperature-controlled stage.

**Dynamic light scattering (DLS)**
DLS measurements were conducted to determine the hydrodynamic diameters of the EF, EL, FL, EFL complexes. PAs were dissolved in Milli-Q water at a concentration of 0.1% (w/v), and ASM was similarly diluted using Milli-Q water. The two or three solutions were mixed at a 1:1 (v/v) and equilibrated for 10 min at room temperature prior to analysis. Size measurements were performed using a Zetasizer instrument (Nano-ZS ZEN 3600, Malvern Instruments).

**Scanning electron microscopy (SEM)**
For SEM, the microstructural morphology of EF, EL, FL, EFL hydrogels, was examined. Samples were fixed with 4% paraformaldehyde (PFA) for 30 min, then dehydrated through a graded ethanol series (70%, 90%, 96%, and 100%). Critical point drying was carried out using an EM CPD300 system (Leica Microsystems, Germany). Samples were sputter-coated with a 10-nm-thick gold layer before imaging. SEM micrographs were acquired using an Inspect Q600 microscope (FEI, The Netherlands).

**Three-dimensional (3D) printing**
3D printing was performed using a commercial extrusion-based bioprinter (INKREDIBLE+, CELLINK, MA, USA). Digital models were created in Autodesk Fusion 360 (Autodesk Inc., San Francisco, CA, USA) and converted into G-code instructions using Slic3r (v 1.3.0; https://slic3r.org/). For printing, a 2 wt% PA-AGD hydrogel precursor was supplemented with thioflavin T (ThT) (0.5 mg/mL) to enable visual monitoring and extruded through a sterile 22-gauge conical nozzle into a support bath composed of ASM. The printhead moved at a constant speed of 40 mm/s during deposition.

**HUVECs cell culture and co-culture with ELK1-FK materials**
Human umbilical vein endothelial cells (hUVECs: Lonza, C2519A), originally expanded in EGM™-2 medium, were cultured in EGM™-2 BulletKit medium (Lonza, CC-3156 & CC-4176) under standard conditions (37 °C, 5% $CO_2$). The medium was refreshed every three days until cells reached 80% confluency. Cells between passages 2 and 4 were used for all experiments. Prior to cell seeding, ELK1–FK filament gels were washed three times with phosphate-buffered saline (PBS) 8 hours after assembly and sterilized under UV light for 45 minutes. Each filament gels was then placed into a well of a 48-well plate, with either the inner or outer surface oriented upward. For both ELK1-FK filament gel and viscous gel, 500 μL of EGM™-2 medium containing 50,000 hUVECs were added to each well. Cells were cultured at 37 °C under 5% $CO_2$ for 24 hours.

**Acknowledgements**
This work was supported by the National Natural Science Foundation of China (NSFC) for the Excellent Young Scientists Fund (Overseas, 0214530013), NSFC for Distinguished Young Scholars (82302837), the China Aerospace Science and Technology Corporation (0231530004), and the Strategic Partnership Research Funding (HUST-Queen Mary University of London, No.2022-HUST-QMUL-SPRF-07). The work was also supported by the ERC Proof-of-Concept grant NOVACHIP, the NIHR Nottingham Biomedical Research Centre at University of Nottingham, Nottingham, UK, and the project AOCMF-21-04S from the AO Foundation, AO CMF. The AO CMF is a clinical division of the AO Foundation — an independent medically guided not-for-profit organization. We thank the Medical Subcentre of Huazhong University of Science and Technology (HUST) Analytical & Testing Centre. We thank Professor Richard OC Oreffo (Bone and Joint Research Group, University of Southampton, UK) for helpful discussions and provision of Laponite® material. JCRC is grateful for funding from the Spanish Government (grant PID2022-137484OB-I00 funded by MCIN/AEI/ 10.13039/501100011033 and by ERDF, EU), from the Department of Education, *Junta de Castilla y León* (grant VA188P23 and CLU-2023-1-05 cofunded by ERDF, EU) and from *Centro en Red de Medicina Regenerativa y Terapia Celular de Castilla y León*.

**Author contributions**
Y.W. and A.M. conceived the project. Y.W. carried out the experiments. Y.W. and A.M. supervised the study. A.v.T. and T.T. conducted computer simulations. J.W. performed the Cryo-TEM. Z.Y. and S.S. performed the energy barrier calculations. J.C.R.C. offered the ELK1 material. L.A. offered the FK and FR materials.

**Competing interests:** The authors declare no competing interests.

**Correspondence and requests for materials**
should be addressed to Yuanhao Wu and Alvaro Mata

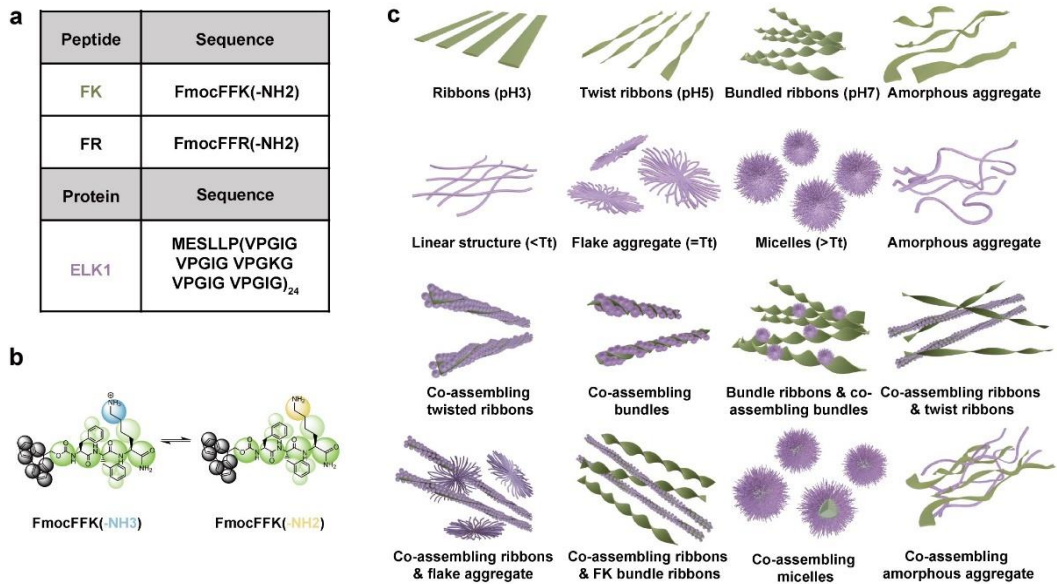

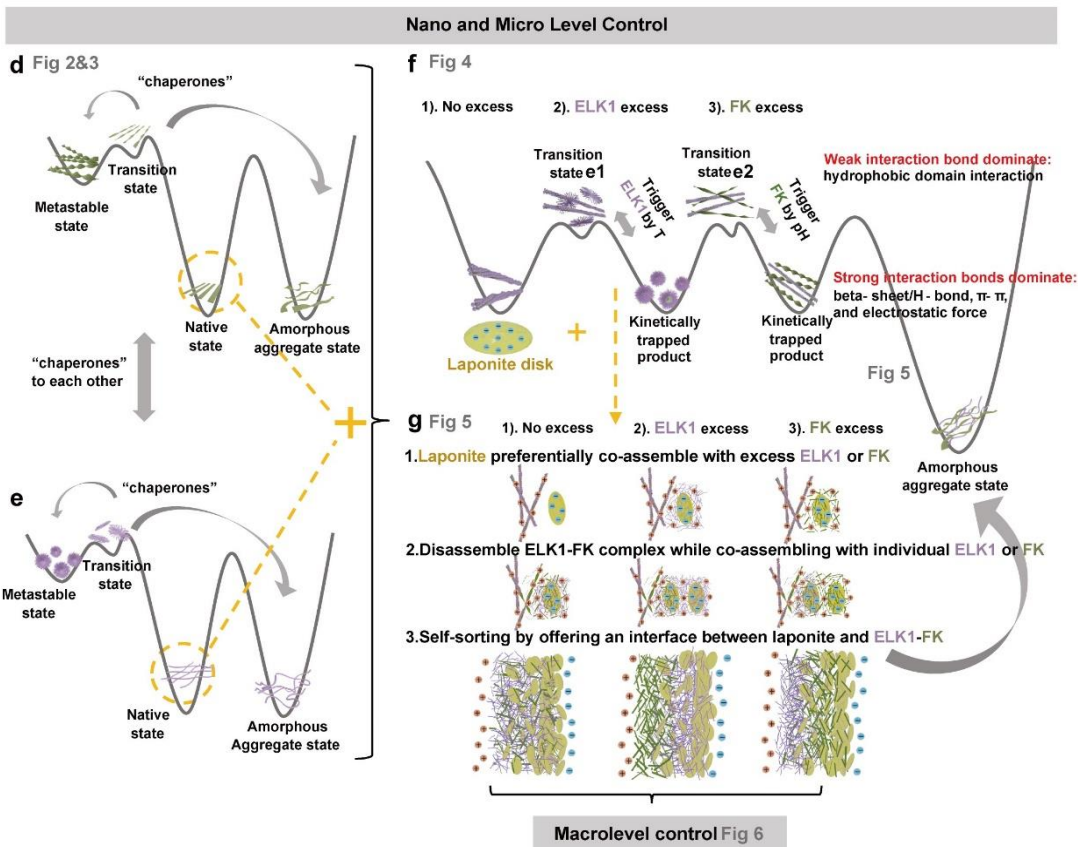

**Figure 1. Schematic representation of the ELK1-FK co-assembly energy landscape.**
(a) Molecular structures of FK, FR, and ELK1. (b) The colour scheme represents the surface (non-grey) and interior of the FmocFFK aggregates (grey) with protonated (blue) and deprotonated (yellow) Lys side chain shown in different colours. (c) Schematic diagrams illustrating self-assembly and co-assembly structures of FK (green) and ELK1 (purple). (d) Energy landscape of FK showing the ribbons-to-bundles transition state at pH 5 (refer to Fig. 2). FK transitions to a nearby energy minimum (metastable state) as bundled ribbons through environmental pH change to pH 7 or interaction with molecular "chaperones" (refer to Fig. 3). The native state and amorphous aggregates of FK represent the energy landscape minima. (e) Energy landscape of ELK1 illustrating the transition from linear assemblies to micelles at its transition temperature (Tt). This transition can be modulated by co-assembly with molecular "chaperones" or by increasing temperature above Tt. FK and ELK1 can act as mutual chaperones (refer to Fig. 3). ELK1's native state and amorphous aggregates represent its energy minima. (f) Energy landscape scenarios of FK-ELK1 co-assembly under three concentration ratios: (1) No excess, (2) ELK1 excess, and (3) FK excess. Co-assembly is driven by weak hydrophobic interactions. Excess ELK1 or FK can transition to nearby energy minima, forming kinetically trapped products with ordered or disordered structures governed by strong intermolecular forces (β-sheet/H-bonds, π–π stacking, electrostatic interactions). Reversible transitions indicate a non-equilibrium state (refer to Fig. 4), while irreversible transitions into amorphous aggregates indicate equilibrium collapse (refer to Fig. 5) (g) Self-sorting at the interface between Laponite® suspension and ELK1-FK solution. Laponite® preferentially co-assemble with excess components, disrupting FK-ELK1 complexes and forming amorphous aggregates at the system's energy minima due to dominant electrostatic interactions (refer to Fig. 5). Accessing transition states (TS) is critical for bridging microscale to macroscale assembly by lowering the energy barrier (refer to Fig. 6).

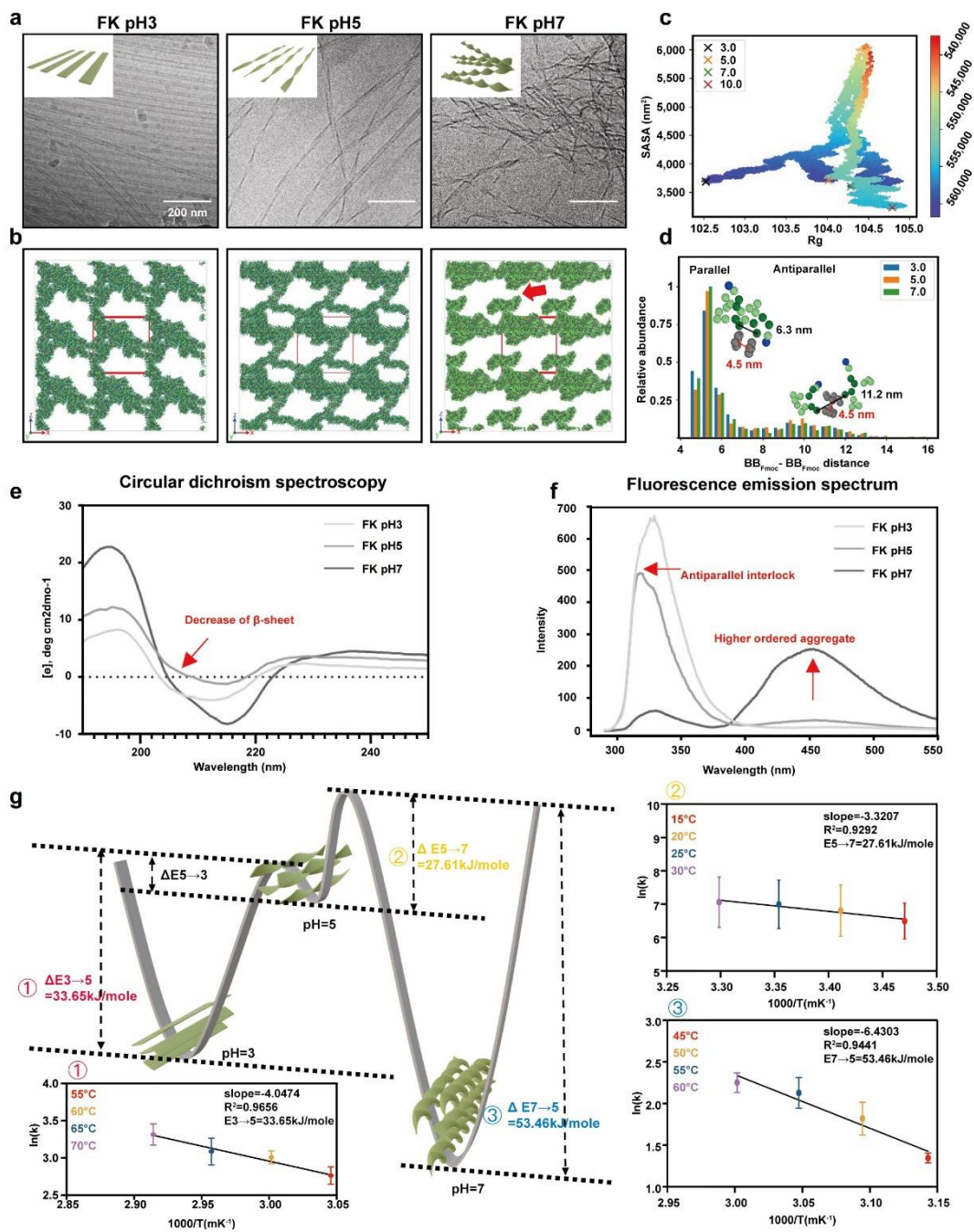

**Figure 2. FK self-assembly and transition state (TS) behavior at pH 5.** (a) Cryo-TEM images of FK self-assembled as ribbons (pH 3), twisted ribbons (pH 5), and bundled ribbons (pH 7). (b) CGMD revealing structural details corresponding to each assembly state. (c) FK at pH 5 exhibits the highest SASA, indicative of a TS. (d) Antiparallel Fmoc-Fmoc stacking at pH 5 confirmed by minimum center-of-geometry distance calculations between FK monomers. (e) CD and (f) fluorescence spectroscopy confirm antiparallel interlock changes in Fmoc-Fmoc stacking at pH 5. (g) Energy landscape of FK at varying pH levels, derived from energy gap calculations between neighboring pHs by DLS data fitted to the Arrhenius equation, confirming FK's TS at pH 5.

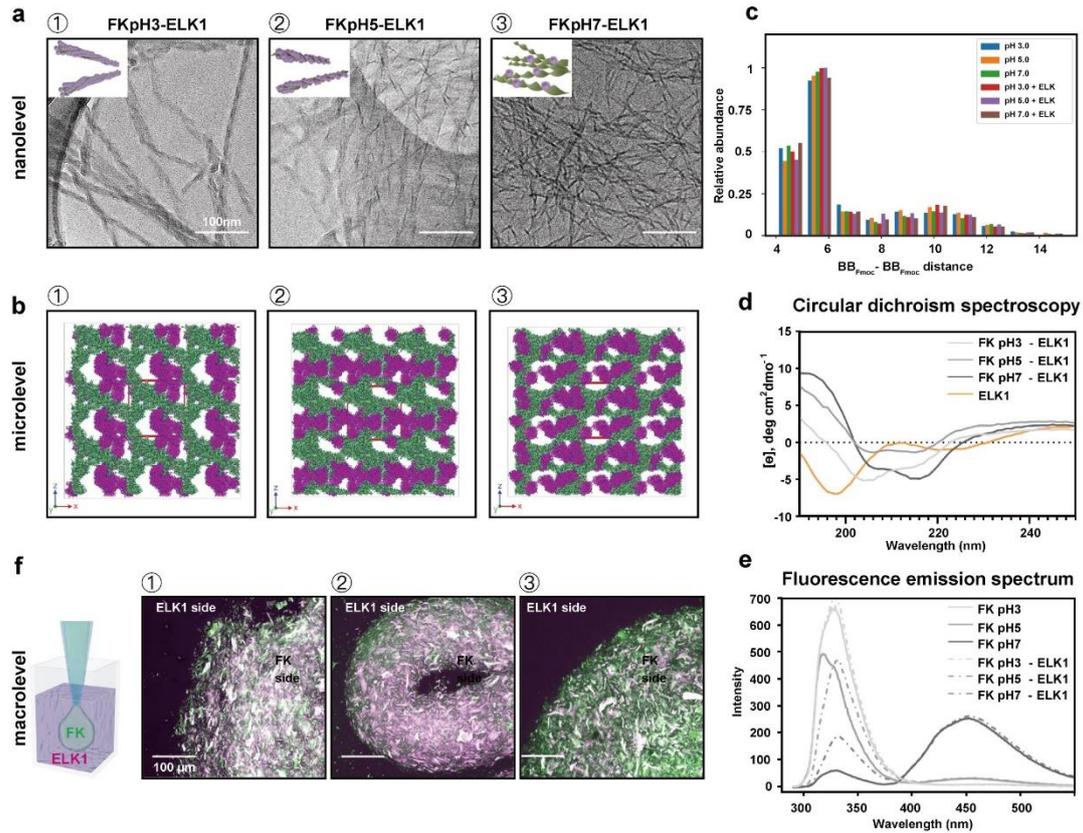

**Figure 3. FK-ELK1 co-assembly at the nanoscale and structural stabilization at FK's TS (pH 5).** (a) Cryo-TEM images of FK-ELK1 co-assembled structures: twisted ribbons (FK pH 3), bundled ribbons (FK pH 5), and combined ribbons and bundles (FK pH 7). (b) CGMD simulations highlight the most robust structure at FK pH 5, attributed to maximal ELK1 adsorption. (c) Fmoc-Fmoc stacking shifts, with reduced antiparallel configurations. (d) CD and (e) fluorescence emission confirm increased α-helix content and decreased antiparallel β-sheet structures upon co-assembly with ELK1. (f) Schematic diagrams illustrating how to assess the LLPS by releasing FK into ELK1 solution. Confocal microscopy visualizes filamentous structures formed when FK solutions at varying pH are introduced into ELK1 solutions, with the most robust structure observed at pH 5 due to rapid co-assembly.

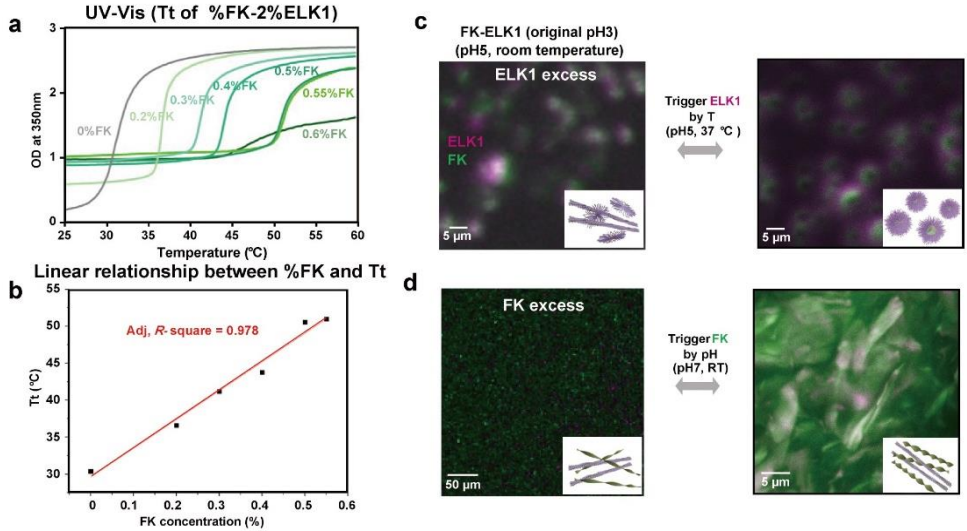

**Figure 4. Microscale tuning of FK-ELK1 co-assembly *via* concentration ratios and transition states.** (a) UV-Vis analysis shows the increase in ELK1-FK system $T_t$ with higher FK concentrations (0–0.6%) at fixed ELK1 (2%). No excess components exist at 0.55% FK; below this, ELK1 self-assembly dominates, and above it, FK self-assembly prevails. (b) Linear correlation between $T_t$ and FK concentration (<0.55%) enables predictive tuning of $T_t$. (c) Design of the co-assembly system with $T_t$ aligned to physiological temperature (37°C). Confocal images reveal reversible micelle structures (purple: ELK1, green: FK) in ELK1-excess systems. (d) FK-excess systems reversibly form co-assembled ribbons and FK bundles upon pH changes.

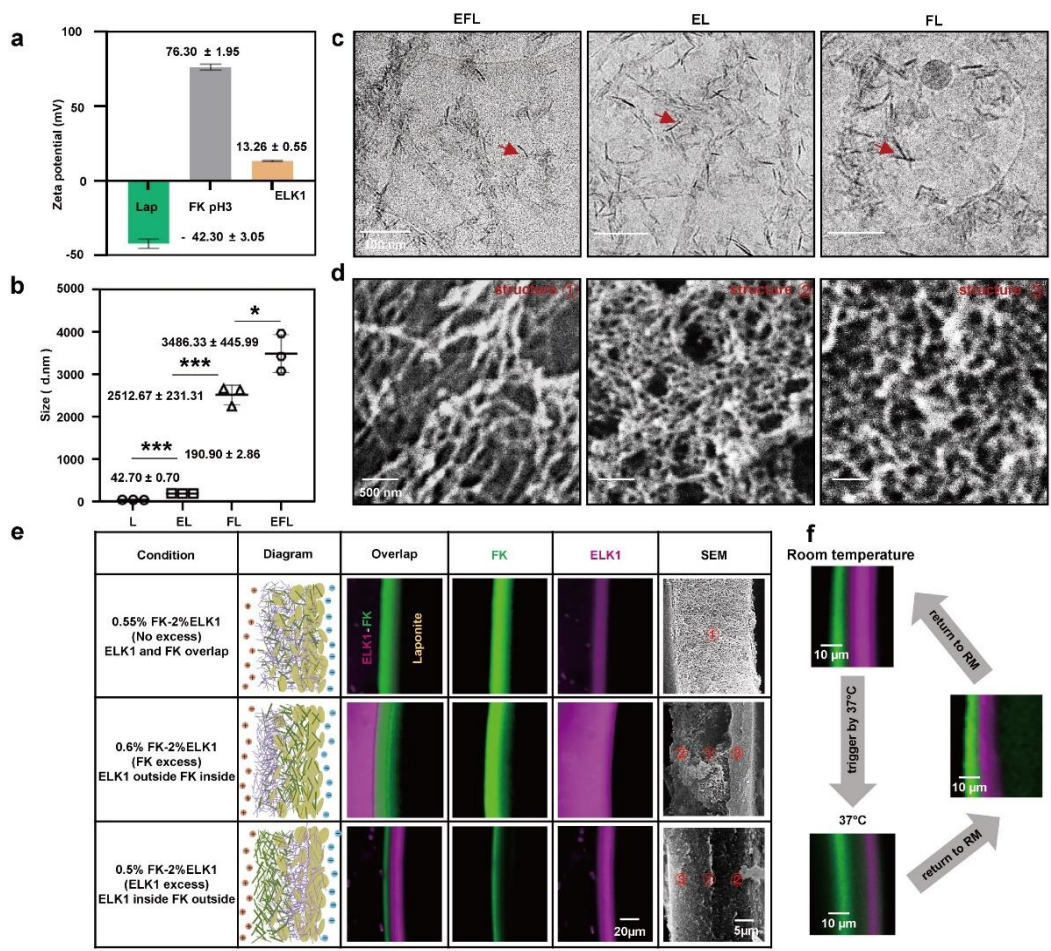

**Figure 5. Self-sorting and non-equilibrium living properties in FK (F)-ELK1 (E) systems *via* Laponite®-induced energy minimization.** (a) Zeta potential analysis confirms negative charges on Laponite® (Lap) and positive charges on FK and ELK1. (b) DLS reveals larger aggregate sizes in EFL systems compared to EL, FL, or Laponite® alone. (c) Cryo-TEM shows disruption of FK and ELK1 ordered structures upon Laponite® introduction (red arrows). (d) SEM images display distinct morphologies in EFL (①), EL (②), and FL (③) systems. (e) Confocal and SEM images confirm self-sorting membrane formation at liquid-liquid interfaces under no excess, FK-excess, and ELK1-excess conditions. (f) ELK1-excess membranes exhibit partial thermal reversibility at designed Tt. Error bars represent ±*s.d. (n = 3); *p < 0.05* (one-way ANOVA).

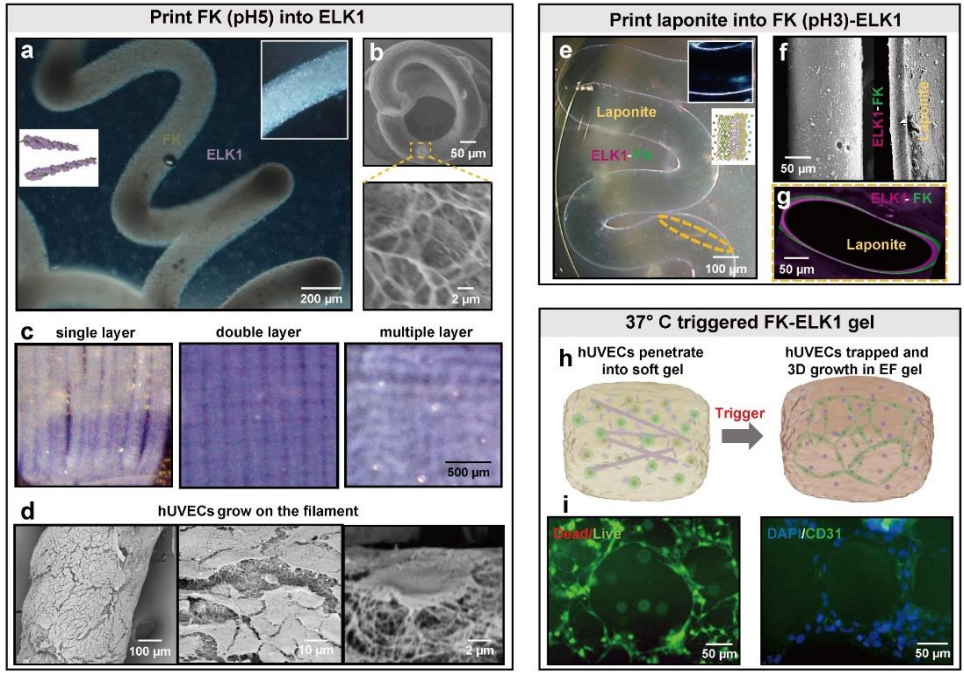

**Figure 6. Supramolecular biofabrication of FK-ELK1 co-assemblies with biofunctional properties.** (a) Stereomicroscopy demonstrates the printability of FK-ELK1 co-assemblies, producing high-fidelity filaments when FK solution at TS is extruded into ELK1 solution. Birefringence (inset) confirms uniform gel-like structure. (b) SEM reveals hierarchical filament structures, supporting (c) multi-layered 3D grid printing and (d) biocompatibility with hUVECs. (e) Tubular structures formed by extruding Laponite® into FK-ELK1 solutions (FK pH 3, ELK1 excess), supported by (f) SEM and (g) confocal imaging of self-sorting membranes. (h) Schematic of an innovative 3D cell culture strategy: FK-ELK1 liquid transitions to micelle gels at 37°C, encapsulating suspended cells. Live/Dead assays confirm this system promotes hUVEC tube formation without growth factors.